\documentclass[a4paper,12pt]{article}
\usepackage{graphicx}
\usepackage{cite}
\usepackage{amssymb,amsmath,latexsym,enumerate}
\usepackage{subfigure,epsfig}
\usepackage{array,longtable,lscape,bigints}

\begin{document}

\title{On the integrability conditions for a family of the Li\'{e}nard--type equations}
\author{Nikolai A. Kudryashov, Dmitry I. Sinelshchikov,\\
\texttt{nakudr@gmail.com}, \texttt{disine@gmail.com}, \vspace{0.5cm}\\
Department of Applied Mathematics, \\ National Research Nuclear University MEPhI, \\ 31 Kashirskoe Shosse, 115409 Moscow, Russian Federation}

\date{}

\maketitle

\begin{abstract}
We study a family of Li\'{e}nard--type equations. Such equations are used for the description of various processes in physics, mechanics and biology and also appear as traveling--wave reductions of some nonlinear partial differential equations. In this work we find new conditions for the integrability of this equations family. To this end we use an approach, which is bases on application of nonlocal transformations. By studying connections between this family of Li\'{e}nard--type equations and type III Painlev\'{e}--Gambier equations, we obtain four new integrability criteria. We illustrate our results by providing examples of some integrable Li\'{e}nard--type equations. We also discuss relationships between linearizability via nonlocal transformations of this family of Li\'{e}nard--type equations and other integrability conditions for this family of equations.

\end{abstract}

\\


\section{Introduction}

Nonlinear ordinary differential equations appear in a vast range of applications \cite{Borisov,Polyanin,Borisov2014,Borisov2014a,Borisov2015}. One of the widely used families of such equations is the family of the Li\'{e}nard--type equations that has the form:
\begin{equation}
y_{zz}+f(y)y_{z}^{2}+g(y)y_{z}+h(y)=0,
\label{eq:L1}
\end{equation}
where $f$, $g$ and $h$ are arbitrary functions. We assume that functions $g(y)$ and $h(y)$ do not vanish, i.e. we consider the dissipative case of equation \eqref{eq:L1}. Notice that at $f(y)=0$ from \eqref{eq:L1} we obtain the classical Li\'{e}nard equation.

Particular cases of equation \eqref{eq:L1} are used for the description of many phenomena in physics and mechanics (see, e.g. \cite{Andronov,Polyanin}). For example, nonlinear oscillators of Van der Pol and Duffing types are members of family of equations \eqref{eq:L1}. Furthermore, traveling--wave reductions of many important for applications nonlinear partial differential equations belong are particular cases of equation \eqref{eq:L1}. For instance, traveling--wave reductions of a nonlinear convection--diffusion equation and the generalized Fisher equation equation are members of family \eqref{eq:L1} (see, e.g. \cite{Polyanin2011,Harko2015}).

Recently, there has been some interest in finding integrable subcases of equation \eqref{eq:L1}. Notice that in this work under integrability we understand a possibility to find the closed--form general solution to a member of family \eqref{eq:L1}. For example, the Prelle--Singer method was applied to find some integrable cases of equation \eqref{eq:L1} \cite{Lakshmanan2005}. Some integrable equations of type \eqref{eq:L1} were found with the help of the Jacobi last multiplier method in \cite{Nucci2008,Guha2009,Nucci2010}. Authors of \cite{Lakshmanan2015} obtained families of equations of type \eqref{eq:L1} which can be integrated by the classical Lie method. Linearization of family \eqref{eq:L1} via nonlocal transformations was also studied in \cite{Meleshko2010,Meleshko2011}.

Here we extend previously obtained results and find new integrable families of equations of type \eqref{eq:L1}. To this aim we use an approach based on application of nonlocal transformations that has recently been proposed and developed in \cite{Kudryashov2015,Kudryashov2016,Kudryashov2016A}. The main idea of this approach is to find a connection given by nonlocal transformations between equation \eqref{eq:L1} and its subcases that have the general closed--form solution. As such subcases of equation \eqref{eq:L1} we consider equations of the Painlev\'{e}--Gambier type (see, e.g. \cite{Ince,Kudryashov_book,Kudryashov2014a}). According to Ince's book \cite{Ince} these equations can be divided into eight types: types I--VIII. In works \cite{Kudryashov2016,Kudryashov2016A} connections between the classical Li\'{e}nard equation and equations the Painlev\'{e}--Gambier equations of type I were studied. Notice that below we show that these results can be easily extended to the case of equation \eqref{eq:L1}. In this work we continue this classification and consider connections between equation \eqref{eq:L1} and equations from the Painlev\'{e}--Gambier classification of type III. In this way we find new families of integrable Li\'{e}nard--type equations. We also discuss some of the previously obtained integrable cases of equation \eqref{eq:L1} and show that they follow from the linearizability via nonlocal transformations of the corresponding Li\'{e}nard--type equations. To the best of our knowledge, our results are new.

The rest of this work is organized as follows. In the next section we give four new integrability conditions for family \eqref{eq:L1} along with condition for the linearizability of \eqref{eq:L1} via nonlocal transformations. We also discuss some relationships between integrability conditions for family \eqref{eq:L1} obtained by various approaches. In section 3 we illustrate results obtained in Section 2 and provide four new examples of integrable Li\'{e}nard--type equations and their general solutions. In the last section we briefly summarize and discuss our results.

\section{Main results}

In this section we use the following transformations
\begin{equation}
w=F(y), \quad d\zeta=G(y)dz, \quad F_{y}G\neq0,
\label{eq:L1_1}
\end{equation}
where $\zeta$ and $w$ are new independent and dependent variables correspondingly, to study connections between equation \eqref{eq:L1} and equations of the Painlev\'{e}--Gambier type. We also discuss linearizability conditions for equation \eqref{eq:L1} via \eqref{eq:L1_1}.

Now we need to determine equations from the Painlev\'{e}--Gambier classification that will be considered in this work. First, here we study connections between \eqref{eq:L1} and type III Painlev\'{e}--Gambier equations. Note that Painlev\'{e}--Gambier equations of types II and IV--VIII should be considered elsewhere. Second, since equation \eqref{eq:L1} is an autonomous equation and transformations \eqref{eq:L1_1} preserve this property we consider only autonomous equations from the Painlev\'{e}--Gambier classification. Third, we omit equations that are nonlocally equivalent to the Painlev\'{e}--Gambier equations of types I and II. As a result, we find ten type III Painlev\'{e}--Gambier equations that satisfy these requirements: they are canonical forms of equations XXIV, XXV, XXVI, XXVII, XXVIII, XXXV, XXXVI and non--canonical form of equations XXIII, XXX, XIX. Note that here and below we use Ince's notation \cite{Ince} for the Painlev\'{e}--Gambier equations.

Equation XXVI is equivalent via transformations \eqref{eq:L1_1} to a non--canonical form of equation XXIII. The same is true for equations XXX and XXIII. Since a combination of transformations \eqref{eq:L1_1} is a transformation of type \eqref{eq:L1_1}, with out loss of generality, we can consider only one of these equations. Equation XXIV can be linearized via \eqref{eq:L1_1}, and, thus, is omitted further. Transformations between equation \eqref{eq:L1} and equations XXXV, XXXVI have a cumbersome form and we do not study them here. Therefore, below we consider connections between equation \eqref{eq:L1} and equations XXIII (non--canonical form), XXV, XXVII and XXVIII.

First of all, we demonstrate that family of equations \eqref{eq:L1} can be normalized with the respect to transformations \eqref{eq:L1_1} and its canonical form is the classical Li\'{e}nard equation. Indeed, the following proposition holds

\textbf{Proposition 1.} Equation \eqref{eq:L1} can be transformed into the classical Li\'{e}nard equation
\begin{equation}
y_{\zeta\zeta}+\tilde{g}(y)y_{\zeta}+\tilde{h}(y)=0,
\label{eq:L1_3}
\end{equation}
with the help of transformations \eqref{eq:L1_1} with $F(y)=y$ and $G(y)=\exp\{-\int f dy\}$, where
$\tilde{g}(y)=\exp\{\int f dy\}g(y)$ and $\tilde{h}(y)=\exp\{2\int f dy\}g(y)$. Since a combination of transformations \eqref{eq:L1_1} is a transformation of type \eqref{eq:L1_1}, equation \eqref{eq:L1} and \eqref{eq:L1_3} are equivalent with respect to \eqref{eq:L1_1}.

As a result, we see that, with out loss of generality, one can assume that $f(y)\equiv0$.  However, we believe that it is more convenient to study equation \eqref{eq:L1} instead of equation \eqref{eq:L1_3}, since in practice we often deal with equations of type \eqref{eq:L1}. To obtain the corresponding results for equation \eqref{eq:L1_3} we can simply assume that $f(y)\equiv0$.

Let us remark that Proposition 1 allows us to extend the integrability conditions obtained in works \cite{Kudryashov2016,Kudryashov2016A} for the classical Li\'{e}nard equation to the case of equation \eqref{eq:L1}. Indeed, applying inverse form of transformations \eqref{eq:L1_1} from Proposition 1 to integrability criteria from works \cite{Kudryashov2016,Kudryashov2016A} we obtain integrability criteria for family \eqref{eq:L1}.

Now let us turn to our main results. Throughout this work we use the following notation
\begin{equation}
\mathcal{G}=\left.\int\right.e^{\int f(y) dy}g(y)dy+\kappa,
\end{equation}
where $\kappa$ is an arbitrary constant. We also denote by $\zeta_{0}$ and $C_{i}$, $i=1,\ldots,6$ arbitrary constants.

We start with a non-canonical form of Painlev\'{e}--Gambier equation XXIII:
\begin{equation}
w_{\zeta\zeta}-\frac{3}{4w}w_{\zeta}^{2}+2\alpha w_{\zeta}+w^{2}+3\alpha^{2}w=0,
\label{eq:Ince_XXIII}
\end{equation}
where $\alpha\neq0$ is an arbitrary parameter.

The general solution of this equation is
\begin{equation}
w=3\alpha^{2}g_{3}e^{-2\alpha(\zeta-\zeta_{0})}\wp^{-2}\{e^{-\alpha(\zeta-\zeta_{0})},0,g_{3}\},
\label{eq:Ince_XXIII_gs}
\end{equation}
where $g_{3}$  is an arbitrary constant. Notice that here and below we denote by $\wp$ the Weierstrass elliptic function.

Conditions for the equivalence between equations \eqref{eq:L1} and \eqref{eq:Ince_XXIII} via \eqref{eq:L1_1} can be expressed as the following statement:

\textbf{Theorem 1.} Equation \eqref{eq:L1} can be transformed into \eqref{eq:Ince_XXIII} by means of transformations \eqref{eq:L1_1} with
\begin{equation}
F=\lambda_{1}\mathcal{G}^{4}, \quad G=\frac{1}{2\alpha}g(y),
\label{eq:Ince_XXIII_nlt}
\end{equation}
if the following correlation holds
\begin{equation}
h=\frac{g(y)}{16\alpha^{2}}\mathcal{G}\left(\lambda_{1}\mathcal{G}^{4}+3\alpha^{2}\right)e^{-\int f dy}.
\label{eq:Ince_XXIII_c}
\end{equation}
\textbf{Proof.} First, we express $y_{z}$ and $y_{zz}$ via $w_{\zeta}$ and $w_{\zeta\zeta}$. Then, we substitute the result into \eqref{eq:L1} and require that the corresponding ordinary differential equation for $w$ is \eqref{eq:Ince_XXIII}. Consequently, we find two ordinary differential equations for $F(y)$ and $G(y)$ and an algebraic relation for functions $f$, $g$ and $h$. Solving these equations we get formulas \eqref{eq:Ince_XXIII_nlt} and \eqref{eq:Ince_XXIII_c}. This completes the proof.

Now we turn to equation XXV of the Painlev\'{e}--Gambier type that has the form
\begin{equation}
w_{\zeta\zeta}-\frac{3}{4w}w_{\zeta}^{2}+\frac{3}{2}w w_{\zeta}+\frac{1}{4}w^{3}-\gamma w-2\delta=0,
\label{eq:Ince_XXV}
\end{equation}
where $\gamma$ and $\delta\neq0$ are arbitrary parameters.

The general solution of this equation can be written as follows
\begin{equation}
w=\frac{2\delta}{2v_{\zeta}+v^{2}-\gamma}, \quad v=\frac{\Psi_{\zeta}}{\Psi},
\label{eq:Ince_XXV_gs}
\end{equation}
where $\Psi$ is the general solution of the equation
\begin{equation}
\Psi_{\zeta\zeta\zeta}-\gamma \Psi_{\zeta}-\delta\Psi=0.
\label{eq:Ince_XXV_gs_1}
\end{equation}
Although it may seem that solution \eqref{eq:Ince_XXV_gs} contains three arbitrary constants, say $C_{1}$, $C_{2}$ and $C_{3}$, without loss of generality, one can set one of them equal to one.

Let us we present the criterion of equivalence between equation \eqref{eq:L1} and \eqref{eq:Ince_XXV}.

\textbf{Theorem 2.} Equation \eqref{eq:L1} can be transformed into \eqref{eq:Ince_XXV} via transformations \eqref{eq:L1_1} with
\begin{equation}
F=\lambda_{2}\mathcal{G}^{4/5}, \quad G=\frac{2g(y)}{3\lambda_{2}}\mathcal{G}^{-4/5},
\end{equation}
if the following correlation holds
\begin{equation}
h=\frac{5g(y)}{36\lambda_{2}^{3}}\left(\lambda_{2}^{3}\mathcal{G}^{12/5}-4\lambda_{2}\gamma \mathcal{G}^{4/5}-8\delta\right)\mathcal{G}^{-7/5}e^{-\int f(y) dy}.
\end{equation}
\textbf{Proof.} The proof is similar to that of Theorem 1 and is omitted.

Now we consider equation XXVII of the Painlev\'{e}--Gambier type. We study a particular case of this equation, which corresponds to $f=\phi=0$ according to Ince's notation, and has the form
\begin{equation}
w_{\zeta\zeta}-\frac{n-1}{n}\frac{w_{\zeta}^{2}}{w}+\frac{n-2}{n}\frac{w_{\zeta}}{w}+\frac{1}{nw}-n\epsilon w=0.
\label{eq:Ince_XXVII}
\end{equation}
Here $n\neq 0$ is an integer number. In the case of $n=-1$ and $n=-2$ equation \eqref{eq:Ince_XXVII} can be linearized via \eqref{eq:L1_1}. The case of $n=1$ was considered in \cite{Kudryashov2016A}, while in the case of $n=2$ equation \eqref{eq:Ince_XXVII} does not belong to family \eqref{eq:L1}. Thus, further, we assume that $n\neq-2,-1,0,1,2$.

The general solution of equation \eqref{eq:Ince_XXVII} has the form
\begin{equation}
w=\psi^{n}\left(C_{4}+\int \psi^{-n}d\zeta\right), \quad \psi=C_{5}e^{\sqrt{\epsilon}\zeta}+C_{6}e^{-\sqrt{\epsilon}\zeta}.
\label{eq:Ince_XXVII_gs}
\end{equation}
Note that in fact solution \eqref{eq:Ince_XXVII_gs} contains only two arbitrary constants and, without loss of generality, one can set either $C_{5}$ or $C_{6}$ equal to $1$.

\textbf{Theorem 3.} Equation \eqref{eq:L1} can be transformed into \eqref{eq:Ince_XXVII} with the help of transformations \eqref{eq:L1_1} with
\begin{equation}
F=\lambda_{3} \mathcal{G}^{-\frac{n}{n-1}}, \quad G=\frac{n\lambda_{3}g(y)}{n-2} \mathcal{G}^{-\frac{n}{n-1}},
\label{eq:Ince_XXVII_nlt}
\end{equation}
if the following correlation holds
\begin{equation}
h=\frac{(n-1)g(y)}{(n-2)^2}\bigg(\epsilon n^{2}\lambda_{3}^{2}\mathcal{G}^{-\frac{2n}{n-1}}-1\bigg)\mathcal{G}\,e^{-\int f dy}.
\label{eq:Ince_XXVII_c}
\end{equation}
\textbf{Proof.} The proof is similar to that of Theorem 1 and, therefore, is omitted.

The final equation of the Painlev\'{e}--Gambier type, which is considered here, is equation XXVIII:
\begin{equation}
w_{\zeta\zeta}-\frac{1}{2w}w_{\zeta}^{2}+ww_{\zeta}-\frac{1}{2}w^{3}+\frac{72\beta}{w}=0,
\label{eq:Ince_XXVIII}
\end{equation}
where $\beta\neq0$ is an arbitrary parameter. Notice that in the case of $\beta=0$ equation \eqref{eq:Ince_XXVIII} can be linearized by transformations \eqref{eq:L1_1}, and, thus, we do not consider this case.

The general solution of \eqref{eq:Ince_XXVIII} is the following
\begin{equation}
w=\frac{6(\wp^{2}\{\zeta-\zeta_{0},12\beta,g_{3}\}-\beta)}{\wp_{z}(\zeta-\zeta_{0},12\beta,g_{3})},
\label{eq:Ince_XXVIII_gs}
\end{equation}
where $g_{3}$ is an arbitrary parameter.

Let us formulate the criterion of equivalence between equations \eqref{eq:L1} and \eqref{eq:Ince_XXVIII} via \eqref{eq:L1_1}:

\textbf{Theorem 4.} Equation \eqref{eq:L1} can be transformed into \eqref{eq:Ince_XXVIII} by means of transformations \eqref{eq:L1_1} with
\begin{equation}
F=\lambda_{4}\mathcal{G}^{2/3}, \quad G=\frac{g(y)}{\lambda_{4}}\mathcal{G}^{-2/3},
\end{equation}
if the following correlation holds
\begin{equation}
h=\frac{3g(y)}{4\lambda_{4}^{4}}\mathcal{G}^{-5/3}\left(144\beta-\lambda_{4}^{4}\mathcal{G}^{8/3}\right)e^{-\int f(y) dy}.
\end{equation}
\textbf{Proof.} The proof is similar to that of Theorem 1 and is omitted.

Now we explicitly obtain linearizability conditions for equation \eqref{eq:L1} via transformations \eqref{eq:L1_1}. The following proposition holds:

\textbf{Proposition 2.} Equation \eqref{eq:L1} can be transformed into the damped harmonic oscillator
\begin{equation}
w_{\zeta\zeta}+\mu w_{\zeta}+\sigma w=0,
\label{eq:dho}
\end{equation}
by means of transformations \eqref{eq:L1_1}, if the following correlation on functions $f$, $g$ and $h$ holds
\begin{equation}
h(y)=\frac{\sigma }{\mu^{2}}\,e^{-\int f(y) dy}\,g(y)\,\mathcal{G}.
\label{eq:dho_c}
\end{equation}
In this case transformations \eqref{eq:L1_1} have the form
\begin{equation}
F(y)=\lambda \mathcal{G},\quad G(y)=\frac{1}{\mu}g(y).
\label{eq:dho_nlt}
\end{equation}
\textbf{Proof.}  The proof is similar to that of Theorem 1 and is omitted.

Let us compare condition \eqref{eq:dho_c} for the linearizability of equation \eqref{eq:L1} with conditions for the integrability of \eqref{eq:L1} obtained with the help of the other methods. If we assume that $g(y)=k_{1}\int \exp\{\int f(y) dy\} dy+k_{2}$, where $k_{1}$ and $k_{2}$ are arbitrary parameters, from \eqref{eq:dho_c} we obtain the condition for the linearizability of equation \eqref{eq:L1} via point transformations, that was found by the Lie method in \cite{Lakshmanan2015}. Actually, this result can be obtained from the linearizability criterion for the classical Li\'{e}nard equation (see, \cite{Kudryashov2016}) since equations \eqref{eq:L1} and \eqref{eq:L1_3} are equivalent due to Proposition 1. If we consider condition for the integrability of \eqref{eq:L1} obtained in \cite{Harko2015} with the help of the Chiellini lemma, we will again see that this condition is equivalent to \eqref{eq:dho_c}. One can also see that condition for the integrability of the classical Li\'{e}nard equation obtained with the help of the Jacobi last multiplier (see \cite{Nucci2010}) also follows from \eqref{eq:dho_c} if we assume that $f(y)=0$. Finally, note that condition \eqref{eq:dho_c} can be obtained from the results of work \cite{Meleshko2010}.

We can also see that criteria for the integrability of \eqref{eq:L1} obtained in Theorems 1--4 do not coincide with both linearizability criterion \eqref{eq:dho_c} and criteria obtained in \cite{Lakshmanan2015} by the Lie method. Therefore, we found four new integrable families of the Li\'{e}nard--type equations. In the next section we illustrate these results by several new examples of the integrable  Li\'{e}nard--type equations.

\section{Examples}

In this section we consider applications of Theorems 1--4 and provide four new examples of the integrable Li\'{e}nard--type equations along with their general solutions. We give one example for each integrability condition obtained in the previous section. Notice that in the examples given below we denote by $a\neq0$ an arbitrary parameter.

\emph{Example 1. Application of Theorem 1.}

\begin{figure}[!h]
\center
\includegraphics[width=0.75\textwidth]{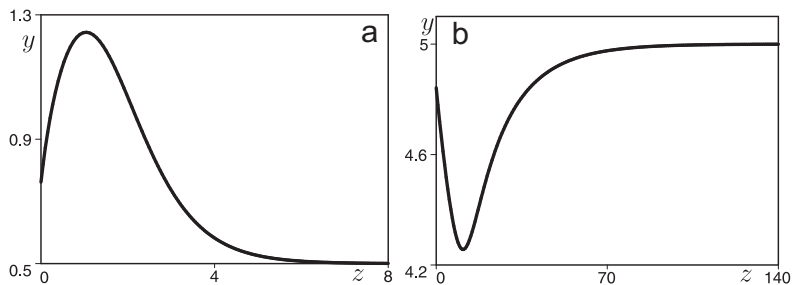}
\caption{Solution \eqref{eq:s1_1} of equation \eqref{eq:s1} at: a) $g_{3}= \lambda=1$, $\kappa=-1$, $\alpha=1/2$, $a=2$ and $\zeta_{0}=2$ (plus sign in \eqref{eq:s1}); b) $g_{3}=10$, $\lambda=1$, $\kappa=-10$, $\alpha=1/2$, $a=2$ and $\zeta_{0}=4/3$ (minus sign in \eqref{eq:s1}).  }
\label{f1}
\end{figure}

Assume that $f(y)=2y/(1+y^{2})$, $g(y)=a/(1+y^{2})$. Then using \eqref{eq:L1} and Theorem 1 we find the corresponding Li\'{e}nard equation
\begin{equation}
y_{zz}+\frac{2y}{1+y^2}y_{z}^{2}+\frac{a}{1+y^{2}} y_{z}+\frac{a(ay+\kappa)(\lambda(ay+\kappa)^{4}+3\alpha^{2})}{16\alpha^{2}(1+y^{2})^{2}}=0.
\label{eq:s1}
\end{equation}
Let us remark that equation \eqref{eq:s1} can be considered as a dissipative generalization of the Mathews-Lakshmanan oscillator \cite{Lakshmanan1974}.
The general solution of \eqref{eq:s1} can be obtained with the help of \eqref{eq:Ince_XXIII_gs} and \eqref{eq:Ince_XXIII_nlt} and is expressed as follows
\begin{equation}
\begin{gathered}
y=\frac{1}{a}\left[\pm \left(\frac{3\alpha^{2}g_{3}e^{-2\alpha(\zeta-\zeta_{0})}}{\lambda}\wp^{-2}\{e^{-\alpha(\zeta-\zeta_{0})},0,g_{3}\}\right)^{1/4}-\kappa\right], \\
z=\left.\int\right. \frac{2\alpha}{a}(1+y^{2}) d\zeta.
\label{eq:s1_1}
\end{gathered}
\end{equation}
Solution \eqref{eq:s1_1} describes various pulse--type structures and damped oscillations. We demonstrate plots of this solution at certain values of the parameters in Fig.\ref{f1}.

\emph{Example 2. Application of Theorem 2.}

\begin{figure}[!h]
\center
\includegraphics[width=0.99\textwidth]{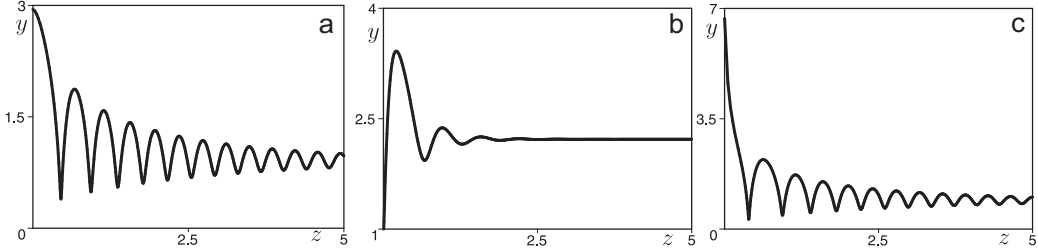}
\caption{Solution \eqref{eq:s3_1} at: a) $\gamma=-10$, $\delta=1/2$, $C_{1}=1$, $C_{2}=-5$ and $C_{3}=-7.7$; b) $\gamma=-20$, $\delta=15$, $C_{1}=-1.5$, $C_{2}=-2$ and $C_{3}=4$; c) $\gamma=-10$, $\delta=1/2$, $C_{1}=0.495$, $C_{2}=0.5$ and $C_{3}=1$.  }
\label{f2}
\end{figure}

Suppose that $f(y)=3/(4y)$, $g(y)=y^{2}$ and $\lambda_{2}=2^{-3/5}15^{-1/5}$. Then, applying Theorem 2 we get the corresponding Li\'{e}nard--type equation
\begin{equation}
y_{zz}+\frac{3}{4y}y_{z}^{2}+y^{2}y_{z}+\frac{1}{27y^{4}}\bigg(y^{9}-225\gamma y^{3}-3375 \delta\bigg)=0.
\label{eq:s3}
\end{equation}
Equation \eqref{eq:s3} can be considered as a stationary reduction of a nonlinear convection--diffusion equation (see, e.g. \cite{Polyanin2011,Harko2015,Kudryashov2016C}).

The general solution of equation \eqref{eq:s3} has the form
\begin{equation}
y=\left(\frac{15}{2}w\right)^{1/3}, \quad z=\frac{1}{5}\int y d\zeta,
\label{eq:s3_1}
\end{equation}
where $w$ is defined by \eqref{eq:Ince_XXV_gs} and \eqref{eq:Ince_XXV_gs_1}. Notice that in order to obtain a real solution we have to use in \eqref{eq:s3_1} a linear combination of real and imaginary parts of the general solution of equation \eqref{eq:Ince_XXV_gs_1}. We demonstrate plots of solution \eqref{eq:s3_1} obtained in this way in Fig. \ref{f2}. We can see that this solution describes various types of damped oscillations.

\emph{Example 3. Application of Theorem 3.}

\begin{figure}[!h]
\center
\includegraphics[width=0.75\textwidth]{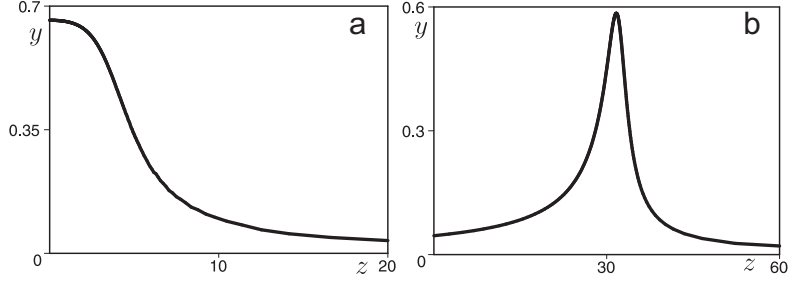}
\caption{Solution \eqref{eq:s5_1} at a) $\epsilon=1$; b) $\epsilon=0.1$.  }
\label{f3}
\end{figure}

We assume that $f(y)=-11/(4y)$, $g(y)=-y$, $\lambda_{3}=3^{-1/3}2^{-1/3}$ and $\kappa=0$. Then, with the help of Theorem 3 we find the corresponding Li\'{e}nard--type equation
\begin{equation}
y_{zz}-\frac{11}{4y}y_{z}^{2}-yy_{z}+y^{3}-\frac{9\epsilon}{4}y^{5}=0.
\label{eq:s5}
\end{equation}
We obtain the general solution of equation \eqref{eq:s5} using \eqref{eq:Ince_XXVII_gs} and \eqref{eq:Ince_XXVII_nlt} as follows
\begin{equation}
\begin{gathered}
y=\frac{8}{3}w, \quad z=\frac{4}{3}\int y^{-2}d\zeta, \\ w=\frac{\psi}{12C_{5}^{2}\sqrt{\epsilon}}\left(12\sqrt{\epsilon}C_{3}C_{5}^{2}\psi^{3}-3e^{-2\sqrt{\epsilon}\zeta}\psi+2C_{6}e^{-3\sqrt{\epsilon}\zeta}\right),\\ \psi=C_{5}e^{\sqrt{\epsilon}\zeta}+C_{6}e^{-\sqrt{\epsilon}\zeta}.
\end{gathered}
\label{eq:s5_1}
\end{equation}
We present plots of solution \eqref{eq:s5_1} at specific values of the parameters in Fig.\ref{f3}. One can see that this solution describes kink and pulse type structures.

\emph{Example 4. Application of Theorem 4.}

\begin{figure}[!h]
\center
\includegraphics[width=0.45\textwidth]{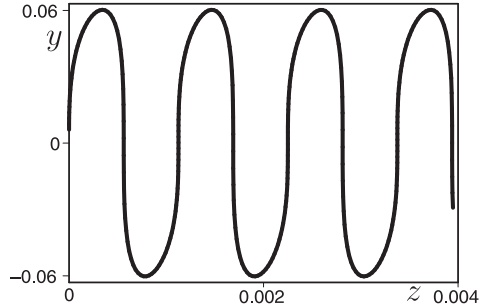}
\caption{Solution \eqref{eq:s7_1} at $\beta=1$, $g_{3}=0.1$, $a=1$ and $\zeta_{0}=10$.  }
\label{f4}
\end{figure}

Let us suppose that $f(y)=1/(2y)$, $g(y)=-a^{3}/y^{3}$, $\kappa=0$ and $\lambda=2^{-2/3}3^{-1/3}$. As a result, with the help of Theorem 4, we find that the following Li\'{e}nard--type equation
\begin{equation}
y_{zz}+\frac{1}{2y}y_{z}^{2}-\frac{a^{3}}{y^{3}}y_{z}+\frac{a^{6}}{2y^{5}}-\frac{5832\beta}{a^{2}y}=0,
\label{eq:s7}
\end{equation}
has the general solution
\begin{equation}
y=\frac{a^{2}\wp_{z}(\zeta-\zeta_{0},12\beta,g_{3})}{18(\wp^{2}\{\zeta-\zeta_{0},12\beta,g_{3}\}-\beta)}, \quad
z=-\frac{1}{3a}\int y^{2}d\zeta.
\label{eq:s7_1}
\end{equation}
Solution \eqref{eq:s7_1} describes periodic oscillations in the case of $1728\beta^{3}>27g_{3}^{2}$. At $1728\beta^{3}=27g_{3}^{2}$ solution \eqref{eq:s7_1} degenerates and is expressed via the trigonometric functions and again is periodic. When $1728\beta^{3}<27g_{3}^{2}$ solution \eqref{eq:s7_1} becomes singular on the real line. In Fig.\ref{f4} we present a plot of solution \eqref{eq:s7_1} for values of the parameters that satisfy $1728\beta^{3}>27g_{3}^{2}$.

In this section we have constructed four examples of integrable Li\'{e}nard--type equations along with their general solutions. We believe that these solutions have been constructed for the first time.

\section{Conclusion}
In this work we have studied family of Li\'{e}nard--type equations \eqref{eq:L1}. We have considered connections, given by nonlocal transformations, between this family of equations and the Painlev\'{e}--Gambier equations of type III. As a consequence, we find four new integrability conditions for family of equations \eqref{eq:L1}. We have shown that corresponding families of Li\'{e}nard--type equations cannot be either linearized via point or nonlocal transformations or integrated by the Lie method. In order to illustrate our results we have presented four new examples of the integrable Li\'{e}nard--type equations and have found their general closed--form solutions. We have also shown that some known integrability conditions for family \eqref{eq:L1} follow from the linearizability via nonlocal transformations of the corresponding Li\'{e}nard--type equations.

\section{Acknowledgments}
This research was partially supported by grant for the state support of young Russian scientists
6624.2016.1 and by RFBR grant 14--01--00498.


\begin{thebibliography}{99}

\bibitem{Borisov} A.V. Borisov, I.S. Mamaev, Moder Methods of Integrable Systems Theory, Institute of Computer Sciences, Moscow-Izevsk, 2003.


\bibitem{Polyanin} V.F. Zaitsev, A.D. Polyanin, Handbook of Exact Solutions for Ordinary Differential Equations, Chapman and Hall/CRC, Boca Raton, 2002.

\bibitem{Borisov2014} A. V. Borisov, N.N. Erdakova, T.B. Ivanova, I.S. Mamaev, The dynamics of a body with an axisymmetric base sliding on a rough plane, Regul. Chaotic Dyn. 19 (2014) 607--634.

\bibitem{Borisov2014a} I.A. Bizyaev, A. V. Borisov, I.S. Mamaev, The dynamics of three vortex sources, Regul. Chaotic Dyn. 19 (2014) 694--701.

\bibitem{Borisov2015} A. V. Borisov, A.A. Kilin, I.S. Mamaev, Dynamics and control of an omniwheel vehicle, Regul. Chaotic Dyn. 20 (2015) 153--172.

\bibitem{Andronov} A.A. Andronov, A.A. Vitt, S.E. Khaikin, Theory of Oscillators, Dover Publications, New York, 2011.

\bibitem{Polyanin2011} A.D. Polyanin, V.F. Zaitsev, Handbook of Nonlinear Partial Differential Equations, Chapman and Hall/CRC, Boca Raton-London-New York, 2011.

\bibitem{Harko2015} T. Harko, M.K. Mak, Exact travelling wave solutions of non-linear reaction-convection-diffusion equations—An Abel equation based approach, J. Math. Phys. 56 (2015) 111501.

\bibitem{Lakshmanan2005} V.K. Chandrasekar, M. Senthilvelan, M. Lakshmanan, On the complete integrability and linearization of certain second-order nonlinear ordinary differential equations, Proc. R. Soc. A Math. Phys. Eng. Sci. 461 (2005) 2451--2476.

\bibitem{Nucci2008} M.C. Nucci, P.G.L. Leach, The Jacobi Last Multiplier and its applications in mechanics, Phys. Scr. 78 (2008) 065011.

\bibitem{Guha2009} A. Ghose Choudhury, P. Guha, B. Khanra, On the Jacobi Last Multiplier, integrating factors and the Lagrangian formulation of differential equations of the Painlev\'{e}–Gambier classification, J. Math. Anal. Appl. 360 (2009) 651--664.

\bibitem{Nucci2010} M.C. Nucci, K.M. Tamizhmani, Lagrangians for dissipative nonlinear oscillators: the method of Jacobi last multiplier, J. Nonlinear Math. Phys. 17 (2010) 167--178.

\bibitem{Lakshmanan2015} A.K. Tiwari, S.N. Pandey, M. Senthilvelan, M. Lakshmanan, Lie point symmetries classification of the mixed Li\'{e}nard-type equation, Nonlinear Dyn. 82 (2015) 1953--1968.

\bibitem{Meleshko2010} W. Nakpim, S.V. Meleshko, Linearization of Second-Order Ordinary Differential Equations by Generalized Sundman Transformations, Symmetry, Integr. Geom. Methods Appl. 6 (2010) 1--11.

\bibitem{Meleshko2011} S. Moyo, S.V. Meleshko, Application of the generalised Sundman transformation to the linearisation of two second-order ordinary differential equations, J. Nonlinear Math. Phys. 18 (2011) 213--236.

\bibitem{Kudryashov2015} N.A. Kudryashov, D.I. Sinelshchikov, On the connection of the quadratic Li\'{e}nard equation with an equation for the elliptic functions, Regul. Chaotic Dyn. 20 (2015) 486--496.

\bibitem{Kudryashov2016} N.A. Kudryashov, D.I. Sinelshchikov, On the criteria for integrability of the Li\'{e}nard equation, Appl. Math. Lett. 57 (2016) 114--120.

\bibitem{Kudryashov2016A} N.A. Kudryashov, D.I. Sinelshchikov, On connections of the Li\'{e}nard equation with some equations of Painlev\'{e}--Gambier type, Submitted to J. Math. Anal. Appl.

\bibitem{Ince} E.L. Ince, Ordinary differential equations, Dover, New York, 1956.

\bibitem{Kudryashov_book} N.A. Kudryashov, Methods of nonlinear mathematical physics, Publisher
hous Intellekt, Moscow, 2010.


\bibitem{Kudryashov2014a} A.V. Borisov, N.A. Kudryashov, Paul Painlev\'{e} and his contribution to science, Regul. Chaotic Dyn. 19 (2014) 1--19.

\bibitem{Lakshmanan1974} P.M. Mathews, M. Lakshmanan, On a unique nonlinear oscillator, Q. Appl. Math. 32 (1974) 215–218.

\bibitem{Kudryashov2016C} N.A. Kudryashov, D.I. Sinelshchikov, Analytical solutions of a nonlinear convection-diffusion equation with polynomial sources, Model. Anal. Inf. Syst. 23 (2016) 309--316.

\end{thebibliography}
\end{document}